\documentclass[aoas,nameyear,seceqn,dvips]{arximspdf}
\usepackage{dcolumn}
\usepackage{graphicx}

\doi{10.1214/09-AOAS279}
\volume{4}
\issue{1}
\pubyear{2010}
\firstpage{396}
\lastpage{421}

\makeatletter
\newcolumntype{d}[1]{D{.}{.}{#1}}
\DeclareMathAlphabet\mathcaligr{OMS}{cmsy}{m}{n}
\newcommand{\cal}{\mathcaligr}
\renewcommand{\cite}{\citet}
\renewcommand{\citep}[1]{[\cite{#1}]}

\newcommand{\xall}{\mathbf{x}_i}
\newcommand{\xchosen}{\mathbf{x}_i^{(c)}}
\newcommand{\xprop}{\mathbf{x}_i^{(p)}}
\newcommand{\xother}{\mathbf{x}_i^{(o)}}
\newcommand{\ychosen}{\mathbf{y}_j^{(c)}}
\newcommand{\yprop}{\mathbf{y}_j^{(p)}}
\newcommand{\yother}{\mathbf{y}_j^{(o)}}
\newcommand{\labels}{\mathbf{l}_i}

\makeatother

\begin{document}
\begin{frontmatter}

\title{Variable selection and updating in model-based discriminant
analysis for high dimensional data~with~food authenticity applications}
\runtitle{Variable selection and updating in discriminant analysis}

\begin{aug}
\author[A]{\fnms{Thomas Brendan} \snm{Murphy}\corref{}\ead[label=e1]{brendan.murphy@ucd.ie}\thanksref{t1,t3}},
\author[B]{\fnms{Nema} \snm{Dean}\thanksref{t3}\ead[label=e2]{nema@stats.gla.ac.uk}}\break
\and
\author[C]{\fnms{Adrian~E.}~\snm{Raftery}\ead[label=e3]{raftery@stat.washington.edu}\thanksref{t2,t3}}
\thankstext{t1}{Supported in part by the Science
Foundation of Ireland Basic Research Grant 04/BR/M0057 and
Research Frontiers Programme Grant 2007/RFP/MATF281.}
\thankstext{t2}{Supported in part by NICHD Grant R01 HD054511 and NSF
Grant ATM 0724721.}
\thankstext{t3}{Supported in part by NIH Grant 8 R01 EB002137-02.}
\runauthor{T. B. Murphy, N. Dean and A. E. Raftery}
\affiliation{University College Dublin, University of Glasgow and
University~of~Washington,~Seattle}
\address[A]{
T. B. Murphy\\
School of Mathematical Sciences\\
University College Dublin\\
Belfield, Dublin 4\\
Ireland\\
\printead{e1}} 
\address[B]{
N. Dean\\
Department of Statistics\\
University of Glasgow\\
Glasgow, G12 8QQ\\
United Kingdom\\
\printead{e2}}
\address[C]{
A. E. Raftery\\
Department of Statistics\\
University of Washington, Seattle\\
Box 354320\\
Seattle, Washington 98195-4320\\
USA\\
\printead{e3}}
\end{aug}

\received{\smonth{4} \syear{2009}}
\revised{\smonth{8} \syear{2009}}

%
\begin{abstract}
Food authenticity studies are concerned with determining if food
samples have been
correctly labeled or not. Discriminant analysis methods are an integral
part of the
methodology for food authentication. Motivated by food authenticity
applications,
a model-based discriminant analysis method that includes
variable selection is presented. The discriminant analysis model is
fitted in a semi-supervised manner using both labeled and
unlabeled data. The method is shown to give excellent classification
performance on several high-dimensional multiclass food authenticity
data sets
with more variables than observations. The variables selected by the
proposed method provide information about which variables are
meaningful for classification purposes. A headlong search strategy
for variable selection is shown to be efficient in terms of
computation and achieves excellent classification performance. In
applications to several food authenticity data sets, our proposed
method outperformed default implementations of Random Forests,
AdaBoost, transductive SVMs and Bayesian Multinomial Regression by substantial
margins.
\end{abstract}

%

\end{frontmatter}

\section{Introduction}

Foods that are expensive are subject to potential fraud where rogue
suppliers may attempt to provide a cheaper inauthentic alternative in
place of the more expensive authentic food. Food authenticity studies
are concerned with assessing the veracity of the labeling of food
samples. Discriminant analysis methods are of prime importance in food
authenticity studies where samples whose authenticity is being assessed
are classified using a discriminant analysis method and the labeling
and classification are compared. Samples determined to have potentially
inaccurate labeling can be sent for further testing to determine if
fraudulent labeling has been used.

Model-based discriminant analysis [\cite{bensmail96}, \cite{fraley02}]
provides a framework for discriminant analysis based on parsimonious
normal mixture models. This approach to discriminant analysis has
been shown to be effective in practice and, being based on a
statistical model, it allows for uncertainty to be treated
appropriately.

In many applications, only a subset of the variables in a
discriminant analysis contain any group membership information and
including variables which have no group information increases the
complexity of the analysis, potentially degrading the classification
performance. Therefore, there is a need for including variable
selection as part of any discriminant analysis procedure. Additionally,
if a subset of variables is found to be important for classification
purposes, then it suggests the potential for collecting a smaller
subset of variables using inexpensive methods rather than the full high
dimensional data.

Variable selection can be completed as a preprocessing step prior to
discriminant analysis (a filtering approach) or as part of the analysis
procedure (a wrapper approach). Completing variable selection prior
to the discriminant analysis can lead to variables that have weak
individual classification performance being omitted from the
subsequent analysis. However, such variables could be important for
classification purposes when jointly considered with others. Hence,
performing variable selection as part of the discriminant analysis
procedure is preferred.

Combining variable selection and linear or quadratic discriminant
analysis has been considered previously in the literature; see \citeauthor{mclachlan92}
[(\citeyear{mclachlan92}), Chapter 12] for a review. Many of these
methods are based on measuring the Mahalanobis distance between
groups before and after the inclusion of a variable into the
discriminant analysis model. In the machine learning literature,
\cite{kohavi97} developed a \textit{wrapper} approach for combining
variable selection in \textit{supervised} learning, of which
discriminant analysis is a special case.

Variable selection is of particular importance in situations where
there are more variables than observations available, that is, large
$p$, small $n$ ($n\ll p$) problems [\cite{west03}]. These situations
arise with increasing frequency in statistical applications,
including genetics, proteomics, image processing and food science.
The two food science applications considered in
Section~\ref{se:datadescrip} involve data sets with many more variables
than observations.

In this paper a version of model-based discriminant analysis is
developed by adapting the model-based clustering with variable
selection method of \cite{raftery06}. This method of discriminant
analysis builds a discriminant rule in a stepwise manner by
considering the inclusion of extra variables into the model and also
considering removing existing variables from the model based on
their importance. The stepwise selection procedure is iterated until
convergence.

A brief review of model-based clustering and discriminant analysis
is given in Section~\ref{se:modelbased}. The underlying model for
model-based clustering with variable selection is reviewed in
Section~\ref{se:variableselection} and this model is extended to
model-based discriminant analysis with variable selection in
Section~\ref{se:DAvariableselection}. In Section~\ref{se:updating}
the fitting of the discriminant analysis model is extended to
incorporate semi-supervised updating using both the labeled and
unlabeled observations [\cite{dean06}] in order to improve the
classification performance.

Search strategies for selecting the variables for inclusion and
exclusion are discussed in Section~\ref{se:search}.
A headlong search strategy is proposed that combines
good classification performance and computational efficiency.
The proposed methodology is applied to the high dimensional data sets
in Section~\ref{se:results} and the methodology and results are
discussed in Section~\ref{se:conclusions}.

\section{Data}\label{se:datadescrip}

\subsection{Food authenticity and near infrared spectroscopy} \label{se:data}

An authentic food is one that is what it claims to be.
Important aspects of food description include its process history,
geographic origin, species/variety and purity. Food producers,
regulators, retailers and consumers need to be assured of the
authenticity of food products.

Food authenticity studies are concerned with establishing whether
foods are authentic or not. Many analytical chemistry techniques are
used in food authenticity studies, including gas chromatography,
mass spectroscopy and vibrational spectroscopic techniques (Raman,
ultraviolet, mid-infrared, near-infrared and visible). All of these
techniques have been shown to be capable of discriminating between
certain sets of similar biological materials. \cite{downey96} and
\cite{reid06}
provide reviews of food authenticity studies with an emphasis on
spectroscopic methods. Near infrared (NIR) spectroscopy provides a
quick and efficient method of collecting data for use in food
authenticity studies
[\cite{downey96}]. It is particularly useful because it requires very
little sample preparation and is nondestructive to the samples being tested.

We consider two food authenticity data sets which consist of combined
visible and
near-infrared spectroscopic measurements from food samples of
different types. The aim of the food authenticity study is to
classify the food samples into known groups. The two studies are
outlined in detail in
Sections~\ref{se:meats} and~\ref{se:oliveoil}:
\begin{itemize}
\item Classifying meats into species (Beef, Chicken, Lamb, Pork,
Turkey).
\item Classifying olive oils into geographic origin (Crete, Peloponese, other).
\end{itemize}

\begin{figure}

\includegraphics{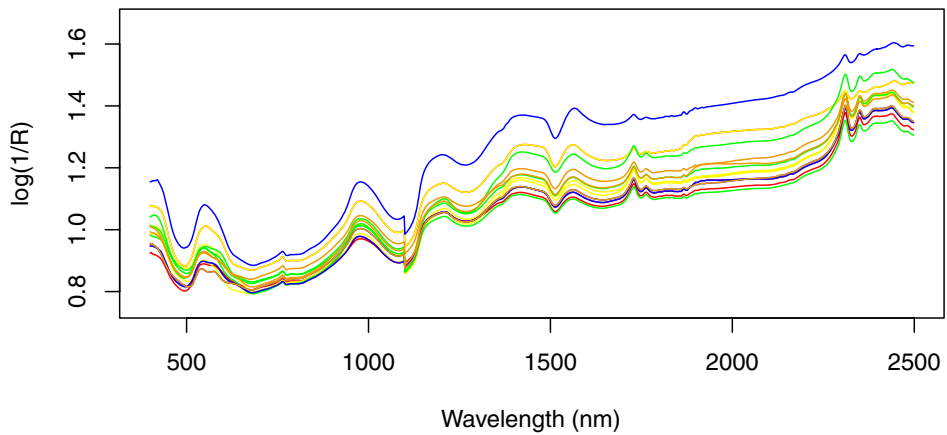}

\caption{The near-infrared spectra recorded for three examples of each
meat species in the study. The discontinuity at 1100~nm is due to
a sensor change at that value. The samples are colored as
Beef${}={}$red, Lamb${}={}$green, Pork${}={}$blue, Turkey${}={}$orange, Chicken${}={}$yellow.}
\label{fi:allmeats}
\end{figure}

In both studies, combined visible and near infrared spectra were
collected in
reflectance mode using an NIRSystems 6500 instrument over the
wavelength range 400--2498 nm at 2~nm intervals. The visible portion
of the spectrum is the range 400--800~nm and the near-infrared
region is the range 800--2498~nm. Hence, the values collected for each
food sample consist of 1050 reflectance values taken at 2~nm intervals
(see, for example, Figure~\ref{fi:allmeats}). For the meat samples,
twenty five separate scans were collected
during a single passage of the spectrophotometer and averaged, after
which the mean spectrum of a reference ceramic tile (16 scans) was
recorded and subtracted from the mean spectrum. A similar process
was used for the olive oil data, but fewer scans were used. Full
details of the spectral data collection process are given in \cite
{mcelhinney99} and \cite{downey03}.

The reflectance values in the visible and near-infrared region are
produced by vibrations in the chemical bonds in the substance being
analyzed. The data are highly correlated due to the presence of a large
number of overlapping broad peaks in this region of the electromagnetic
spectrum and the presence of combinations and overtones. As a result,
even though the data are very highly correlated, the reflectance values
at adjacent wavelengths can have different sources and reflectance
values at very different wavelengths can have the same source. So, the
information encoded in each spectrum is recorded in a complex manner
and spread over a range of locations. \cite{osborne93} provide an
extensive review of the chemical and technological aspects of
near-infrared spectroscopy and its application. Further information on
the combined spectra and their structure is given in Section~\ref
{se:results} where the results of the analysis of the data are given.

Because of the complex nature of the combined spectroscopic data, there
is interest in determining if a small subset of reflectance values
contain as much information for authentication purposes as the whole
spectrum does. If a small number of variables contain sufficient
information for authentication purposes, then this indicates the
possibility of developing portable sensors for food authenticity
studies that are more rapid and have a lower cost than recording the
combined visible and near-infrared spectrum. In fact, portable sensors
have been developed on a commercial basis for the authentication of
Scottish whiskys [\cite{connolly06}] using ultraviolet spectroscopic
technology. Hence, there are motivations for incorporating feature
selection in the classification methods used on these data from the
application and the modeling viewpoints.

The problem of feature selection is especially difficult because the
number of possible subsets of wavelengths that could be selected in
this problem is $2^{1050}$. So, efficient search strategies need to be
used so that a good set of features can be selected without searching
over all possible subsets.

\subsection{Homogenized meat data} \label{se:meats}

\cite{mcelhinney99} constructed a collection of combined visible and
near-infrared spectra from 231 homogenized meat samples in order to
assess the effectiveness of visible and near-infrared spectroscopy as a
tool for determining the correct species of the samples. The samples collected
for this study consist of 55 Chicken, 55 Turkey, 55 Pork, 32~Beef and 34
Lamb samples. The samples were collected over an extended period of
time and from a number of sources in order to ensure a representative
sample of meats.

\begin{figure}[b]

\includegraphics{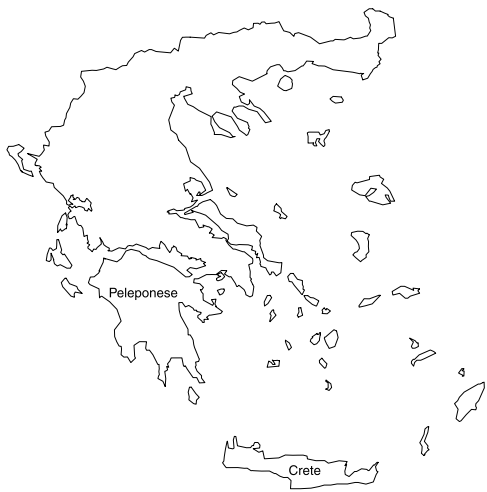}

\caption{Regions of Greece where the olive oil samples were
collected.} \label{fi:greece}
\end{figure}

For each sample, a spectrum consisting of 1050 reflectance measurements was
recorded (as outlined in Section~\ref{se:data}). A plot of all of the
spectra is shown in Figure~\ref{fi:allmeats}. We can see that there is
a discrimination between the red meats (beef and lamb) and the white
meats (chicken, turkey and pork) over some of the visible region
(400--800~nm), but discrimination within meat colors is less clear.

\subsection{Greek olive oils data}
\label{se:oliveoil}

\cite{downey03} recorded near-infrared spectra from a total of 65
extra virgin olive oil samples that were collected from three different
regions in
Greece (18 Crete, 28 Peloponese, 19~other). Each data value consists
of 1050 reflectance values over the visible and near-infrared range.
The aim of their study was to assess the effectiveness of near-infrared
spectroscopy in determining the geographical
origin (see Figure~\ref{fi:greece}) of the oils.

\section{Model-based clustering and discriminant analysis}
\label{se:modelbased}

Model-based clustering
[\cite{banfield93}, \citeauthor{fraley98b} (\citeyear{fraley98b,fraley02}), \cite{McLachlanPeel2000}] uses
mixture models as a framework for cluster analysis. The underlying
model in model-based clustering is a normal mixture model with $G$
components, that is,
\[
f(\mathbf{x})=\sum_{g=1}^{G}\tau_g f(\mathbf{x}|\mu_g,\Sigma_g),
\]
where $f(\cdot|\mu_g,\Sigma_g)$ is a multivariate normal density
with mean $\mu_g$ and covariance~$\Sigma_g$.

A central idea in model-based clustering is the use of constraints
on the group covariance matrices $\Sigma_g$; these constraints use
the eigenvalue decomposition of the covariance matrices to impose
shape restrictions on the groups. The decomposition is of the form
$\Sigma_g=\lambda_g D_g A_g D_g^{T}$, where $\lambda_g$ is the
largest eigenvalue, $D_g$ is an orthonormal matrix of eigenvectors,
and $A_g$ is a diagonal matrix of scaled eigenvalues.
Interpretations for the parameters in the covariance decomposition
are as follows: $\lambda_g{}={}$volume; $A_g{}={}$shape; $D_g{}={}$orientation.
These parameters can be constrained in various ways to be equal or
variable across groups. Additionally, the shape and orientation
matrices can be set equal to the identity matrix.

\cite{bensmail96} developed model-based discriminant analysis
methods using the same covariance decomposition. An extension of
model-based discriminant analysis that allows for updating of the
classification rule using the unlabeled data was developed by
\cite{dean06} and will be described in more detail in
Section~\ref{se:updating}. Model-based clustering and discriminant
analysis can be implemented in the statistics package \textsf{R}
[\cite{R}] using the \texttt{mclust} package
[\citeauthor{fraley99} (\citeyear{fraley99,fraley03,fraley07})].

\subsection{Model-based clustering with variable selection}
\label{se:variableselection}

We argue that variable selection needs to be part of the
discriminant analysis procedure, because completing variable
selection prior to discriminant analysis may lose important grouping
information. This argument is supported by the result of \cite{chang83},
who showed that the principal components corresponding to the
larger eigenvalues do not necessarily contain information about
group structure. This suggests that the commonly used filter approach
of selecting the first few principal components to explain a minimum
percentage of variation can be suboptimal. A similar argument can be
made that selecting discriminating variables without reference to
the grouping variable may miss important variables. In addition,
some variables may contain strong group information when used in
combination with other variables, but not on their own.
Another critique of completing a variable (or feature) selection step
before supervised learning (filtering) is given by \cite{kohavi97}, Section~2.4.

\cite{raftery06} developed a stepwise variable selection wrapper for
model-based
clustering. With their method,
variables are selected in a stepwise manner. Their method involves
the stages:
\begin{itemize}
\item
A variable is proposed for addition to the set of selected clustering
variables. The Bayesian Information Criterion (BIC) is used to compare
a model in which the variable contains extra information about the
clustering beyond the information in the already selected variables
versus a model where the variable doesn't contain additional
information about the clustering beyond the information in the already
selected variables. The variable with the greatest positive BIC
difference is added to the model. If the proposed variable has a
negative BIC difference, then no variable is added.
\item BIC is used to consider whether a variable should be removed from
the model; This step is the reverse of the variable addition step. If
all of the selected variables contain clustering information, then none
is removed from the set of selected clustering variables.
\end{itemize}
This process is iterated until no further variables are added or
removed. This approach, that combines variable selection and cluster
analysis, avoids the problems of completing variable selection
independently of the clustering. While the stepwise variable selection
wrapper proposed in \cite{raftery06} and other wrapper approaches can
give excellent clustering results, there is a considerable
computational burden with wrapper approaches when compared to filtering
approaches; this is because the model needs to be fitted each time a
variable is added or removed from the set of selected clustering variables.

\subsection{Model-based discriminant analysis with variable selection}
\label{se:DAvariableselection}

We adapt the ideas of \cite{raftery06} to produce a discriminant
analysis technique that includes a stepwise variable selection wrapper. This
discriminant analysis method uses a stepwise variable selection
procedure to find a subset of variables that gives good
classification results.

Each stage of the algorithm involves two steps:
\begin{itemize}
\item Determine if a variable (not already selected)
would contribute to the discriminant analysis model. In order to do
this, a model comparison using BIC is used to compare a discriminant
analysis model where the variable contains group information beyond the
information in the already selected variables versus a model where the
variable does not contain group information beyond the information in
the already selected variables. Variables where the BIC difference is
positive are candidates for addition to the set of selected variables;
the procedure for
searching for variables to add to the model is given in
Section~\ref{se:search}.
\item
Determine if any selected variables should be removed from the
discriminant analysis model. This step is the reverse of the variable
addition step. Variables where the BIC model comparison suggests that
the variable does not contain group information are candidates for
removing from the set of selected variables; the procedure for
searching for variables to remove from the model is outlined in
Section~\ref{se:search}.
\end{itemize}

Let $(\mathbf{x}_1,\mathbf{x}_2,\ldots,\mathbf{x}_n)$ be the
observed data values and let
$(\mathbf{l}_1,\mathbf{l}_2,\ldots,\mathbf{l}_n)$ be the group
indicator variables for these observations where $l_{ig}=1$ if
observation $i$ belongs to group $g$ and $l_{ig}=0$ \vspace*{-2pt}otherwise.

Suppose that the observation $\xall$ is partitioned into three
parts: $\xchosen$\vspace*{-1pt} are the variables already chosen; $\xprop$ is
the variable being proposed; $\xother$ are the remaining variables.
The decision on whether to include or exclude a proposed variable
is based on the comparison of two models:
\begin{itemize}
\item Grouping:\vspace*{1.5pt}
$p(\xall|\labels)=p(\xchosen,\xprop,\xother|\labels)=p(\xother
|\xprop,\xchosen)p(\xprop,\xchosen|\labels)$.
\item No Grouping:
$p(\xall|\labels)=p(\xchosen,\xprop,\xother|\labels)=p(\xother
|\xprop,\xchosen)p(\xprop|\xchosen)\times\break p(\xchosen|\labels)$.
\end{itemize}

\begin{figure}[b]

\includegraphics{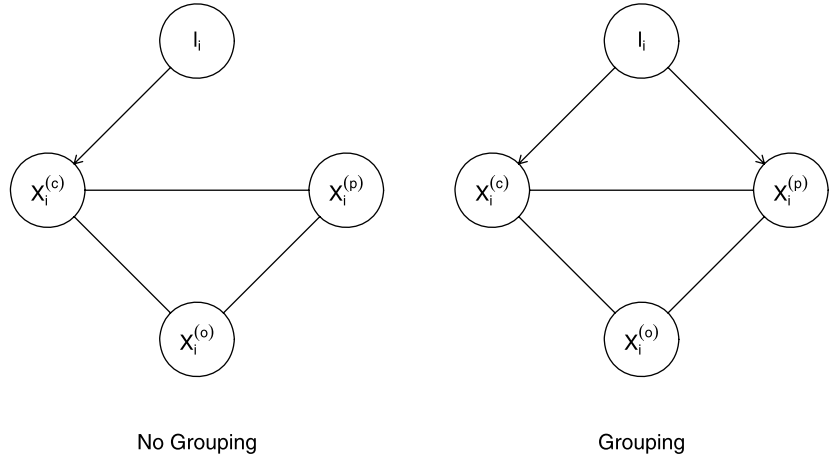}

\caption{A graphical model representation of the Grouping and the
No Grouping models.} \label{fi:graphical}
\end{figure}

Figure~\ref{fi:graphical} shows the difference between the
``Grouping'' and
``No Grouping'' models for $\xall$. If the Grouping model holds,
$\xprop$ provides information about which group the data value belongs to
beyond that provided by $\xchosen$, while if the No Grouping model holds,
$\xprop$ provides no extra information.

\begin{table}
\caption{Constrained covariance structures in model-based
clustering as implemented in~the~\texttt{mclust}~package~for~\textsf{R}} \label{ta:covariance}
\begin{tabular*}{\textwidth}{@{\extracolsep{\fill}}lcccc@{}}
\hline
\textbf{Model ID} & \textbf{Volume} & \textbf{Shape} & \textbf{Orientation} &
\textbf{Covariance ($\bolds\Sigma_{\bolds g}$)}\\
\hline
EII & Equal & Equal spherical & NA & $\lambda I$\\
VII & Variable & Equal spherical & NA & $\lambda_g I$\\
EEI & Equal & Equal & Axis aligned & $\lambda A$\\
VEI & Variable & Equal & Axis aligned & $\lambda_g A$\\
EVI & Equal & Variable & Axis aligned & $\lambda A_g$\\
VVI & Variable & Variable & Axis aligned & $\lambda_g A_g$\\
EEE & Equal & Equal & Equal & $\lambda D A D^T$\\
EEV & Equal & Equal & Variable & $\lambda D_g A D_g^T$\\
VEV & Variable & Equal & Variable & $\lambda_g D_g A D_g^T$\\
VVV & Variable & Variable & Variable & $\lambda_g D_g A_g
D_g^T$\\
\hline
\end{tabular*}
\end{table}

The Grouping and No Grouping models are specified as follows:
\begin{itemize}
\item{Grouping:} We let $p(\xprop,\xchosen|\labels)$ be a normal
density with parsimonious covariance structure as described in
Table~\ref{ta:covariance}. That is,
\begin{eqnarray*}
\bigl(\xprop,\xchosen\bigr)|(l_{ig}=1)
&\sim&
 N\bigl(\mu_{g}^{(p,c)},\Sigma
^{(p,c)}_g\bigr),\\
\labels
&\sim&
\operatorname{Multinomial}(1,\tau),
\end{eqnarray*}
where $\tau=(\tau_1,\tau_2,\ldots,\tau_G)$.
\item{No Grouping:} We let $p(\xchosen|\labels)$ be a normal
density with parsimonious\vspace*{-1pt} covariance structure. In addition,
$p(\xprop|\xchosen)$ is assumed to have a linear regression model
structure. That is,
\begin{eqnarray*}
\xchosen|(l_{ig}=1)
&\sim&
N\bigl(\mu_g^{(c)},\Sigma^{(c)}_g\bigr),\\
\labels
&\sim& \operatorname{Multinomial}(1,\tau),\\
\xprop|\xchosen
&\sim& N\bigl(\alpha+\beta^T\xchosen,\sigma^2\bigr),
\end{eqnarray*}
where $\tau=(\tau_1,\tau_2,\ldots,\tau_G)$.
\end{itemize}

The same model structure is assumed for $p(\xother|\xchosen,\xprop)$
in the Grouping model as in the No Grouping model.
Therefore, this part of the model does not influence the
choice to include $\xprop$ in the model or not.

The decision as to whether the Grouping or No Grouping model is
appropriate is made using the BIC approximation of the log Bayes factor.
The logarithm of the Bayes factor is
%
\begin{equation}
\log (\mbox{Bayes Factor}) = \log\frac{p(\xall|{\cal
M}_G)}{p(\xall|{\cal M}_{NG})}, \label{eq:logbayesfactor}
\end{equation}
where ${\cal M}_G$ is the Grouping model, ${\cal M}_{NG}$ is the No
Grouping model and
\[
p(\xall|{\cal M}_k)=\int p(\xall|\theta_k,{\cal
M}_k)p(\theta_k|{\cal M}_k)\,d\theta_k
\]
is the integrated likelihood of model ${\cal M}_k$.
We use the BIC approximation of the integrated likelihood
in the form
\[
\operatorname{BIC} = 2\times \mbox{log maximized likelihood} - d \log(n),
\]
where $d$ is the number of parameters in the model and $n$ is the
sample size
[\cite{Schwarz1978}].
Following \cite{raftery06},
the log Bayes factor (\ref{eq:logbayesfactor}) can be reduced to
%
\begin{eqnarray}
\log (\mbox{Bayes Factor})
& = &
\log\frac{p(\xprop,\xchosen|{\cal M}_{G})}{p(\xprop|\xchosen
,{\cal M}_{NG})p(\xchosen|{\cal M}_{NG})}
\nonumber\\[-8pt]\\[-8pt]
&\approx&
\frac{1}{2} [\operatorname{BIC}(\mbox{Grouping}) - \operatorname{BIC}(\mbox{No
Grouping})],
\nonumber
\label{eq:logbayesfactor2}
\end{eqnarray}
which only involves $(\xchosen,\xprop)$ and not $\xother$. Variables
with a positive difference in $\operatorname{BIC}(\mbox{Grouping}) - \operatorname{BIC}(\mbox{No
Grouping})$ are
candidates for being added to the model.

At each variable addition stage, the BIC of the grouping model is
calculated using each of the ten covariance structures given in
Table~\ref{ta:covariance} and the model with the highest BIC is
selected for the Grouping model for model comparison purposes.

At each stage, we also check if an already chosen variable should be
removed from the model. This decision is made on the basis of the BIC
difference\vspace*{1pt} in a similar way to previously. In this case, $\xprop$
takes the role of the\vspace*{-1pt} variable to be dropped, $\xchosen$ takes the
role of the remaining chosen variables and $\xother$ are the other
variables. The variables with a positive difference in $\operatorname{BIC}(\mbox{Grouping})
- \operatorname{BIC}(\mbox{No Grouping})$ are candidates for removal from the model; in
this case, the BIC for the no grouping models are computed for all
covariance structures from Table~\ref{ta:covariance} and the model
with the highest BIC is selected as the No Grouping model.


\subsection{Discriminant analysis with updating}
\label{se:updating}

In standard discriminant analysis, the unlabeled data are not used in
the model fitting procedure. However, these data contain information
that is potentially important, especially when very few labeled data
values are available. We can model both the labeled and unlabeled
data as coming from the same model, but where the unlabeled data is
missing the labeling variable, this leads to a mixture model for the
unlabeled data. Hence, the unlabeled data can then be used to help
fit a model to the data. This idea has been investigated by many
authors, including \cite{ganesalingam78} and \cite{oneill78} and more
recently by \cite{dean06}, \cite{chapelle06}, \cite{toher07} and
\cite{liang07}.

Let
$(\mathbf{x}_1,\mathbf{l}_1),(\mathbf{x}_2,\mathbf{l}_2),\ldots
,(\mathbf{x}_N,\mathbf{l}_N)$
be the labeled data and
$\mathbf{y}_1,\mathbf{y}_2,\ldots,\mathbf{y}_M$ be the unlabeled
data. We let $\underline{\mathbf{z}}=(\mathbf{z}_{1},
\mathbf{z}_{2},\ldots, \mathbf{z}_{M})$ be the unobserved (missing)
labels for the unlabeled data.
In this framework, the Grouping and No Grouping models for the
observed data are of the form:

\begin{itemize}
\item{Grouping:} We let $p(\xprop,\xchosen|\labels)$ be a normal
density with parsimonious covariance structure as described in
Table~\ref{ta:covariance}, namely,
\begin{eqnarray*}
\bigl(\xprop,\xchosen\bigr)|(l_{ig}=1)&\sim& N\bigl(\mu_{g}^{(p,c)},\Sigma
^{(p,c)}_g\bigr),\\
\labels&\sim& \operatorname{Multinomial}(1,\tau) .
\end{eqnarray*}
Also, $p(\yprop,\ychosen)$ is a mixture of normals with parsimonious
covariance structures, namely,
\[
\bigl(\yprop,\ychosen\bigr)\sim\sum_{g=1}^{G}\tau_g
N\bigl(\mu_{g}^{(p,c)},\Sigma^{(p,c)}_g\bigr).
\]

\item{No Grouping:} We let $p(\xchosen|\labels)$ be a normal
density with parsimonious covariance structure, namely,
\begin{eqnarray*}
\xchosen|(l_{ig}=1) &\sim& N\bigl(\mu_g^{(c)},\Sigma^{(c)}_g\bigr),\\
\labels&\sim& \operatorname{Multinomial}(1,\tau).
\end{eqnarray*}
We also let $p(\ychosen)$ be a mixture of normal densities with
parsimonious covariance structure, namely,
\[
\ychosen\sim\sum_{g=1}^{G}\tau_g N\bigl(\mu_g^{(c)},\Sigma^{(c)}_g\bigr).
\]
In addition, we assume a linear regression model for\vspace*{-1pt}
$p(\xprop|\xchosen)$ and $p(\yprop|\break\ychosen)$, namely,
\[
\xprop|\xchosen\sim N\bigl(\alpha+\beta^T\xchosen,\sigma^2\bigr)
\]
and
\[
\yprop|\ychosen\sim N\bigl(\alpha+\beta^T\ychosen,\sigma^2\bigr).
\]
\end{itemize}

In both models, we assume an identical model structure\vspace*{-1pt} for
$p(\xother|\xchosen,\xprop)$ and $p(\yother|\ychosen,\yprop)$,
and this doesn't affect the choice to include a variable\vspace*{1pt} in the model
or not.

This model can be fitted using the EM algorithm [\cite{dempster77}]
by introducing the missing labels $\underline{\mathbf{z}}$ into the
model. The calculations involved in fitting the model including the
labeled and unlabeled data follow those outlined in \cite{dean06}.
The maximum likelihood estimates for the
regression part of the model correspond to least squares estimates of
the regression parameters.

The final estimates of the posterior probability of group memberships
produced by the EM algorithm are used to classify the unlabeled observations.
Thus, each observation $j$ is classified into the group $g$ that maximizes
$\hat{z}_{jg}$ over $g$, where
\[
\hat{z}_{jg}=
\frac{\hat{\tau}_g
p(\ychosen|\hat{\mu}^{(c)}_g,\hat{\Sigma}^{(c)}_g)}{\sum
_{g'=1}^{G}\hat{\tau}_{g'}
p(\ychosen|\hat{\mu}^{(c)}_{g'},\hat{\Sigma}^{(c)}_{g'})},
\]
$\ychosen$ is the set of chosen variables, and
$\{(\hat{\tau}_g,\hat{\mu}^{(c)}_g,\hat{\Sigma
}^{(c)}_g)\dvtx g=1,2,\ldots,G\}$
are the maximum likelihood estimates for the unknown model
parameters for this set of chosen variables.

%

\subsubsection{Example}
\label{se:example}

\begin{figure}

\includegraphics{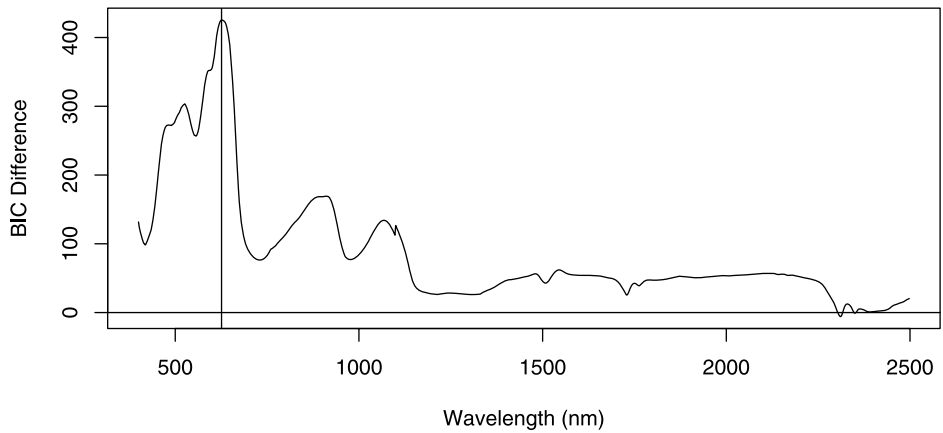}

\caption{A plot of the BIC difference for each wavelength. The
wavelength with the greatest BIC difference is 626~nm.}
\label{fi:meatchosen1}
\end{figure}

\begin{figure}[b]

\includegraphics{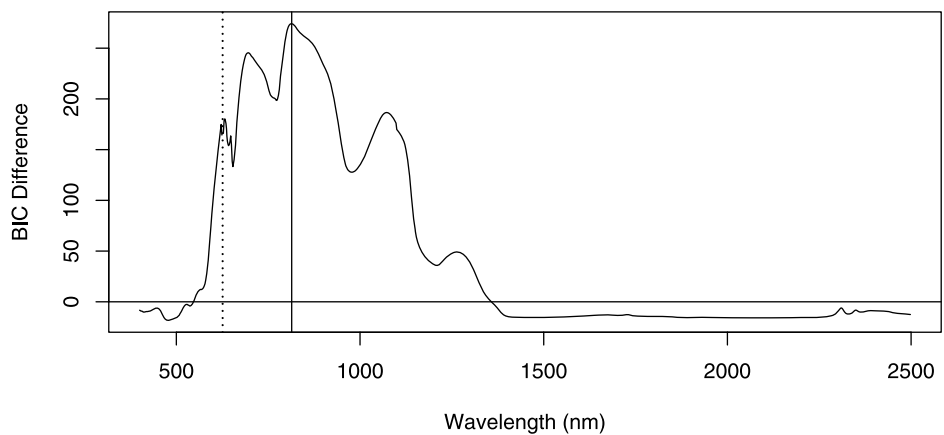}

\caption{A plot of the BIC difference for each wavelength given that
wavelength 626~nm is already accepted. The wavelength with the
greatest BIC difference is 814~nm. Note that wavelengths close to
626~nm still have positive BIC difference values.}\label{fi:meatchosen2}
\end{figure}

\begin{figure}

\includegraphics{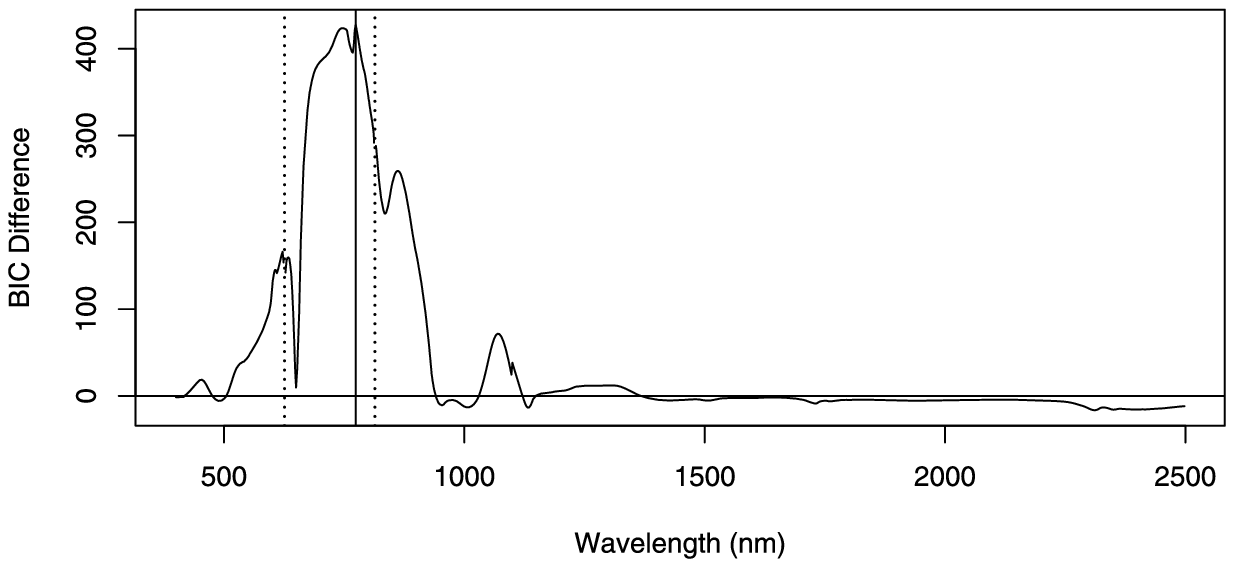}

\caption{A plot of the BIC difference for each wavelength given that
the first two wavelengths chosen (626~nm and 814~nm) are already
accepted. The
wavelength with the greatest BIC difference is
774~nm.}\label{fi:meatchosen3}
\end{figure}

An illustrative example of the BIC calculations when the proposed
algorithm is applied to the meat spectroscopy data is shown in
Figures~\ref{fi:meatchosen1}--\ref{fi:meatchosen3}; half the data of
each type were randomly selected as training data in this example.

The variable selection algorithm begins by selecting 626~nm as the
wavelength with the greatest difference between the Grouping and No
Grouping models (Figure~\ref{fi:meatchosen1}) and the E covariance
structure was chosen. It is worth noting that wavelengths
close to 626~nm still have strong evidence of grouping even though
the spectra are smoothly varying. This phenomenon is due to the fact
that the spectrum consists of a number of overlapping peaks and the
reflectances at adjacent locations can have different sources. As a
result, extra grouping information can be available at wavelengths that
are very close.

Subsequently, the
814~nm wavelength is added to the model
(Figure~\ref{fi:meatchosen2}) and the EEV covariance structure was
chosen. At the third stage, the 774~nm
wavelength is selected (Figure~\ref{fi:meatchosen3}) and the VEV
covariance structure was chosen. The procedure
continues until thirteen wavelengths are selected (details of the
iterations are given in Table \ref{ta:variableselectupdate}) and the
VEV covariance structure is chosen at all subsequent stages.
%
\begin{table}
\caption{A full example of the variable selection procedure used to
classify the meat samples into five types.
The updating procedure was used in this example}
\label{ta:variableselectupdate}
\begin{tabular*}{\textwidth}{@{\extracolsep{\fill}}l@{\,}cd{3.1}ccd{4.1}c@{}}
\hline
\textbf{Iteration} & \textbf{Proposal} & \multicolumn{1}{c}{\textbf{BIC diff.}}
& \textbf{Decision} & \textbf{Proposal} & \multicolumn{1}{c}{\textbf{BIC diff.}} &
\textbf{Decision}\\
\hline
\phantom{1}1 & Add 626~nm & 425.4 & Accepted & & & \\
\phantom{1}2 & Add 814~nm & 274.1 & Accepted & & &\\
\phantom{1}3 & Add 774~nm & 427.4 & Accepted & Remove 774~nm & -427.4 & Rejected\\
\phantom{1}4 & Add 664~nm & 142.6 & Accepted & Remove 626~nm & -120.1 & Rejected\\
\phantom{1}5 & Add 680~nm & 220.1 & Accepted & Remove 774~nm & -78.8 & Rejected\\
\phantom{1}6 & Add 864~nm & 165.2 & Accepted & Remove 774~nm & -91.7 & Rejected\\
\phantom{1}7 & Add 602~nm & 118.9 & Accepted & Remove 774~nm & -26.3 & Rejected\\
\phantom{1}8 & Add 794~nm & 118.3 & Accepted & Remove 774~nm & -86.2 & Rejected\\
\phantom{1}9 & Add 702~nm & 178.6 & Accepted & Remove 774~nm & -127.5 & Rejected\\
10 & Add 1996~nm & 127.5 & Accepted & Remove 1996~nm & -127.5 &
Rejected\\
11 & Add 644~nm & 76.6 & Accepted & Remove 644~nm & -76.6 & Rejected\\
12 & Add 2316~nm & 24.1 & Accepted & Remove 2316~nm & -24.1 & Rejected\\
13 & Add 2310~nm & 103.2 & Accepted & Remove 702~nm & -26.1 & Rejected\\
14 & Add 1936~nm & 10.8 & Accepted & Remove 702~nm & 4.4 & Accepted\\
15 & Add 704~nm & -3.7 & Rejected & Remove 1936~nm & -41.3 & Rejected\\
\hline
\end{tabular*}
\end{table}

Interestingly, many of the chosen wavelengths are in the visible
range (400--800~nm) of the spectrum, indicating that color is
important when separating the meat samples. The closest two wavelengths
that were selected were 2310~nm and 2316~nm and a number of wavelengths
that were selected are approximately 20~nm apart. In summary, the
selected wavelengths are spread out mainly in the visible region, but
some wavelengths were selected in the near-infrared region.

\subsection{Headlong model search strategy} \label{se:search}

The variable selection algorithm dem\-onstrated in
Section~\ref{se:example} is a greedy search strategy. At the
variable addition stages of the algorithm, the variable with the
greatest BIC difference is added and at variable removal stages, the
variable with the greatest BIC difference is removed. The process of
finding the variable with the greatest BIC difference involves
calculating the BIC difference for all variables under
consideration; for the spectroscopic data there are typically just under
1050~variables under consideration at the variable addition stages.
Hence, this search strategy is computationally demanding; this feature
is shared by other wrapper variable selection methods too.

A less computationally expensive alternative is to use a headlong
search strategy [\cite{badsberg92}]. The variable added or removed in
the headlong search strategy need not be the best in terms of having
the greatest BIC difference; it merely needs to be the first
variable considered whose difference is greater than some
pre-specified value (here $\mathit{min.evidence}$). We found that
$\mathit{min.evidence}=0$ gave good results for the applications in this
paper. The headlong strategy has close connections to the
``first-improvement'' moves used in local search algorithms [e.g.,
\cite{hoos05}, Chapter 2.1]. This means that instead of adding the variable
with the greatest evidence for Grouping versus No Grouping,
the first variable found to have a certain amount of evidence for
Grouping versus No Grouping would be added. At the variable
addition stages of the algorithm, the remaining variables are
examined in turn from an ordered list. The initial order of the list
is based on the variables' original BIC differences at the
univariate addition stage; this ordering was used in a similar
context in \cite{yeung05}. We experimented with the
initial ordering and also tried using increasing wavelength and
decreasing wavelength. The classification performance was not
sensitive to the initial ordering, but the selected variables did
depend on the ordering. In the context of increasing and decreasing
wavelength, there was a bias toward selecting low and high
wavelengths, respectively.

Here is a summary of the algorithm:
\begin{enumerate}
\item Select the first variable that is added to be the one that has
the most evidence for Grouping versus No Grouping in terms of greatest
BIC difference
(the same as the first step of the greedy search algorithm).
Create a list of the remaining variables in decreasing order of BIC differences.
\item Select the second variable that is added to be the first variable
in the list of remaining variables with BIC difference for Grouping
versus No Grouping, including the first variable selected, greater than
$\mathit{min.evidence}$. Any variable checked and not used at this stage is
placed at the end of the list of remaining variables.
\item Select the next variable that is added to be the first variable
in the list of remaining variables with BIC difference for Grouping
versus No Grouping,
including the previous variables selected, greater than $\mathit{min.evidence}$.
If no variable has BIC difference greater than $\mathit{min.evidence}$, then no
variable is added at this stage. Any variable checked and not used at
this stage is placed, in turn, at the end of the list of remaining variables.
\item Check in turn each variable currently selected (in reverse order
of inclusion) for evidence of No Grouping (versus Grouping), including
the other selected variables, and remove the first variable with BIC
difference greater than $\mathit{min.evidence}$. If no variable has BIC
difference greater than $\mathit{min.evidence}$, then no variable is removed at
this stage. The removed variable is placed at the end of the list of
other remaining variables.
\item Iterate steps 3 and 4 until two consecutive steps have been
rejected, then stop.
\end{enumerate}

\section{Results}
\label{se:results}

The proposed methodology was applied to the two food authenticity
data sets outlined in Section~\ref{se:data}. In each case, the data
were split so that 50\% of the data were used as labeled data and
50\% as unlabeled. The methodology was applied to 50 random splits
of labeled and unlabeled data and the mean and standard deviation of
the classification rate were computed.

The results were compared to
previously reported performance results for these data and
several widely used alternative techniques: Random Forests [\cite{breiman01}],
AdaBoost [\cite{freund97}], Bayesian Multinomial Regression
[\cite{madigan05}], and Transductive Support Vector Machines
[\cite{vapnik95}, \cite{joachims99}, \cite{collobert06}].

We used the default settings in the \texttt{R} [\cite{R}]
implementations of Random Forests (\texttt{randomForest} version 4.5-30) [\cite
{liaw02}] and AdaBoost (\texttt{adabag} version 1.1)
[\cite{cortes07}]. The use of various parameter settings was explored,
but the results did not vary to a large extent with respect to the
choice of parameter values. For Bayesian Multinomial Regression we used
cross validation to choose between the choice of prior variance
values $\{10^{p}\dvtx p=-4,-3,-2,-1,0,1,2,3,4\}$ as suggested in
\cite{genkin05}. For the Transductive SVM analysis we used the
UniverSVM software version 1.1 [\cite{sinz07}] with a linear kernel
and parameters $(c,s,z)=(100,-0.3,0.1)$; other parameter values were
considered, but the values reported yielded the best classifications.

\subsection{Meats data}
\label{se:meatsresults}

The results achieved on the homogenized meat data
(Section~\ref{se:meats}) are reported in Table~\ref{ta:meat5res}.
These results show that the variable selection and updating method
gives comparable or better performance than previous analyses of
these data; an improved classification rate has been achieved
relative to those achieved by \cite{mcelhinney99} who used factorial
discriminant analysis (FDA), $k$-nearest neighbors ($k$NN), discriminant
partial least squares regression (PLS) and soft independent modeling
of class analogy (SIMCA). Furthermore, a comparable classification
performance has been achieved relative to \cite{dean06} who used
model-based discriminant analysis with updating on a reduced form of
the data derived from wavelet thresholding. The variable selection
and updating procedure gave substantially better performance than
other competing methods for classification.

\begin{table}
\caption{Classification performance on the Meats data for the
variable selection algorithm with updating and for previous analyses
of these data. Mean classification performance for the 50
random~splits~of~the~data~are reported with standard deviations in
parentheses}\label{ta:meat5res}
\begin{tabular*}{\textwidth}{@{\extracolsep{\fill}}ld{2.9}@{}}
\hline
\textbf{Method} & \multicolumn{1}{c@{}}{\textbf{Misclassification rate}}\\
\hline
Variable selection and updating & 6.1\%\ (3.5)\\
Variable selection (greedy) and updating & 5.1\%\ (1.9)\\
Variable selection only& 9.3\%\ (3.6)\\
\cite{dean06} & 5.6\%\ (2.0)\\
\cite{mcelhinney99} & 7.3\%\mbox{--}13.9\%\\
Transductive SVMs & 42.6\%\ (5.7)\\
Random Forests & 20.1\%\ (3.8)\\
AdaBoost.M1 & 20.3\%\ (4.8)\\
Bayesian Multinomial Regression & 34.2\%\ (5.8)\\
\hline
\end{tabular*}
\end{table}

An examination of the misclassification table
(Table~\ref{ta:meat5class}) for the variable selection and updating
method shows that many of the misclassifications were due to the
difficulty in separating the chicken and turkey groups.
Interestingly, no misclassifications were made between the red and
white meats.

\begin{table}[b]
\tablewidth=315pt
\caption{Average classification results for the different meat types
for the variable selection and~updating~classification
method}\label{ta:meat5class}
\begin{tabular*}{315pt}{@{\extracolsep{\fill}}ld{2.1}d{2.1}c@{\hspace*{-15pt}}cd{2.1}d{2.1}@{}}
\hline
&\multicolumn{6}{c@{}}{\textit{\textbf{Predicted}}} \\[-5pt]
&\multicolumn{6}{c@{}}{\hrulefill} \\
\textit{\textbf{Truth}} & \multicolumn{1}{c}{\textbf{Beef}} &
\multicolumn{1}{c}{\textbf{Lamb}}&
& \textbf{Pork} & \multicolumn{1}{c}{\textbf{Turkey}} & \multicolumn{1}{c@{}}{\textbf{Chicken}}\\
\hline
Beef & 98.6& 1.4 && \phantom{9}0.0 & 0.0& 0.0\\
Lamb & 1.4& 98.6 && \phantom{9}0.0 & 0.0& 0.0\\[3pt]
Pork & 0.0& 0.0 && 99.2 & 0.5& 0.3\\
Turkey& 0.0& 0.0 && \phantom{9}0.0 &88.2& 11.8\\
Chicken&0.0& 0.0 && \phantom{9}0.0& 11.1& 88.9\\
\hline
\end{tabular*}
\end{table}

\begin{figure}

\includegraphics{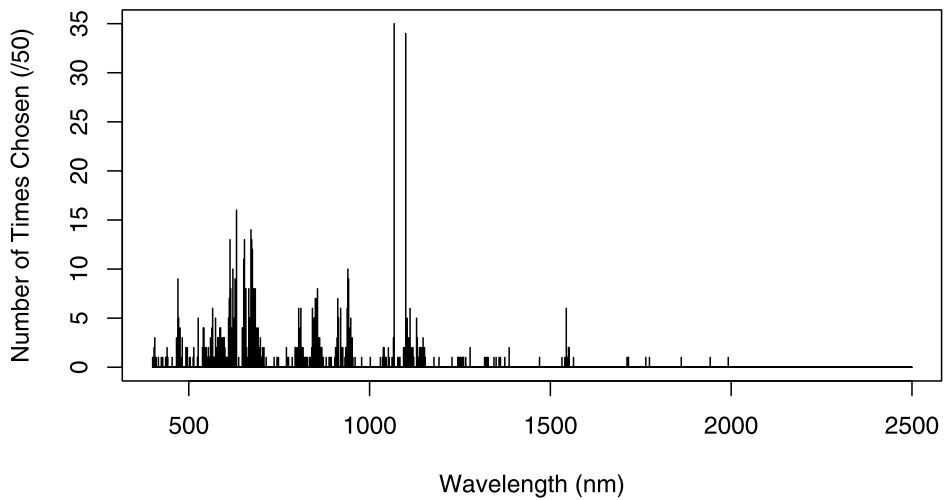}

\caption{Wavelengths chosen in the five meat classification problem
for the variable selection and updating method. The height of the
bars shows how many times the wavelength was chosen in~50~random
splits of the data.} \label{fi:meats5chosen}
\end{figure}

The chosen wavelengths show us which parts of the spectrum are of
importance when classifying samples into different species. We
recorded the chosen wavelengths for each of the 50 sets of results
and these are shown in Figure~\ref{fi:meats5chosen}. We can see that
a large proportion (51\%) of the chosen wavelengths are in the
visible region (400--800~nm), but some regions in the
near-infrared spectrum are also chosen. \citeauthor{liu00} [(\citeyear{liu00}), Table 1]
assign many of the spectral features in the visible part of the
spectrum to different forms of myoglobin such as deoxymyoglobin
(430, 440, 445~nm), oxymyoglobin (545, 560, 575, 585~nm),
metmyoglobin (485, 495, 500, 505~nm) and sulfmyoglobin (635~nm).
Sulfmyoglobin is a product of the reaction of myoglobin with $H_2S$
generated by bacteria, and \cite{arnalds04} found the region of the
spectrum close to 635~nm to be important when separating the red and
white meat samples. The peak at 1100~nm is the wavelength where the
sensor changes in the near-infrared spectrometer and the peak at
1068~nm can be attributed to third overtones of C-H stretch mode and
C-H combination bonds from meat constituents other than oxymyoglobin
[\cite{liu00b}]. The near infrared region consisting of wavelengths
near 1510~nm has been attributed to protein, and a cluster of chosen
wavelengths is close to this region. In all cases, between 13 and 19
wavelengths were chosen for classification purposes.

Following \cite{mcelhinney99} and \cite{dean06}, we combined the
chicken and turkey groups into a poultry group to determine how well
we can classify the homogenized meat samples into four types. The
classification results are reported in Table~\ref{ta:meats4res} and
the misclassifications from the variable selection method with
updating are shown in Table~\ref{ta:meats4class}. There is a
significant improvement in classification performance from all of
the methods. Again, the white and red meats are separated with zero
error.

\begin{table}
\caption{Classification performance on the Meats data for the
variable selection algorithm with updating and for previous analyses
of these data after combining the chicken and turkey into
a~poultry~group.~Mean~classification performance for the 50 random splits
of~the~data~are~reported~with~standard deviations in parentheses}
\label{ta:meats4res}
\begin{tabular*}{\textwidth}{@{\extracolsep{\fill}}ld{2.8}@{}}\hline
\textbf{Method} & \multicolumn{1}{c@{}}{\textbf{Misclassification rate}}\\
\hline
Variable selection and updating & 0.8\%\ (1.3)\\
Variable selection (greedy) and updating& 0.7\%\ (0.7)\\
Variable selection only & 1.8\%\ (3.2)\\
\cite{dean06} & 1.0\%\ (0.9)\\
\cite{mcelhinney99} & 2.6\%\mbox{--}4.3\%\\
Transductive SVMs & 20.9\%\ (8.0)\\
Random Forests & 10.5\%\ (3.3)\\
AdaBoost.M1 & 14.7\%\ (3.7)\\
Bayesian Multinomial Regression & 17.2\%\ (4.9)\\ \hline
\end{tabular*}
\end{table}

\begin{table}[b]
\tablewidth=270pt
\caption{Average classification results for the different meat types
after combining the chicken and turkey into a poultry group. The
results shown are~for~the~variable~selection and updating method}
\label{ta:meats4class}
\begin{tabular*}{270pt}{@{\extracolsep{\fill}}ld{2.1}d{2.1}c@{\hspace*{-15pt}}cd{3.1}@{}}
\hline
&\multicolumn{5}{c@{}}{\textit{\textbf{Predicted}}} \\[-5pt]
&\multicolumn{5}{c@{}}{\hrulefill} \\
\textit{\textbf{Truth}} & \multicolumn{1}{c}{\textbf{Beef}} &
\multicolumn{1}{c}{\textbf{Lamb}}&
& \textbf{Pork} & \multicolumn{1}{c@{}}{\textbf{Poultry}}\\
\hline
Beef & 98.2& 1.8 & & \phantom{9}0.0 & 0.0 \\
Lamb & 2.7 & 97.3 & & \phantom{9}0.0 & 0.0 \\[3pt]
Pork & 0.0 & 0.0 &  &99.1& 0.9 \\
Poultry&0.0& 0.0 & & \phantom{9}0.0 & 100.0 \\
\hline
\end{tabular*}
\end{table}

\begin{figure}

\includegraphics{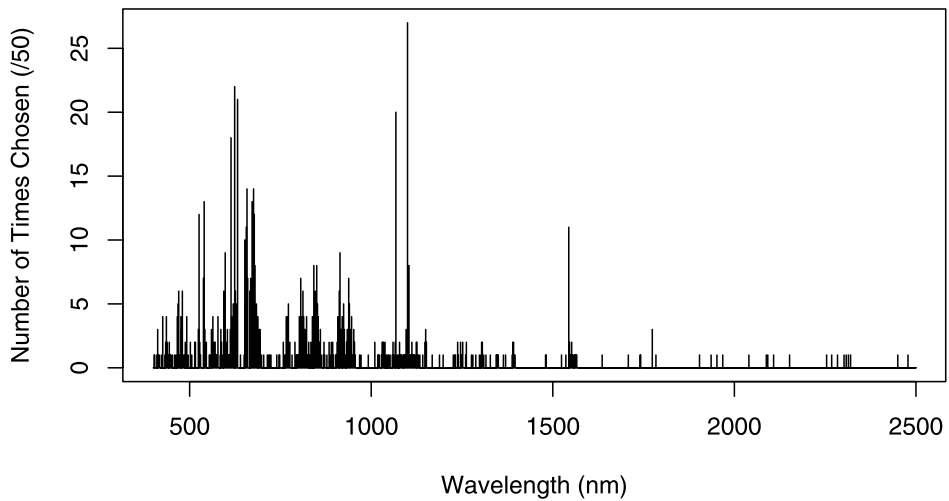}

\caption{Wavelengths chosen in the four meat classification problem
for the variable selection and updating
method.}\label{fi:meats4chosen}
\end{figure}

The wavelengths chosen for the four group classification problem
(Figure~\ref{fi:meats4chosen}) still have a substantial proportion
chosen from the visible part of the spectrum (52\%). In this
application, between 13 and 21 wavelengths were chosen for
classification purposes. The VEV covariance structure was chosen in
almost every run as the final model for both the four and five group
meat classification problems.

\subsection{Greek olive oil data}

The methods were applied to the Greek olive oil data
(Section~\ref{se:oliveoil}), with 50\% of the data being treated as
training data and 50\% as test data. Fifty random splits of training
and test data were used. The misclassification rates achieved on
these data are reported in Table~\ref{ta:oliveclassres}. Variable
selection and updating provides one of the best classification rates
for these data. \cite{downey03} did report a better
misclassification rate (6.1\%) using factorial discriminant analysis
(FDA), but the choice of a subset of wavelengths, data preprocessing
method and classification method (from partial least squares,
factorial discriminant analysis and $k$-nearest neighbors) was made
with reference to the test data classification performance. In
contrast, our model selection was done without any reference to the test
data classification performance.

\begin{table}[b]
\caption{Classification performance on the Olive Oil data for the
variable selection algorithm with updating and for previous analyses
of these data. Mean classification performance for the 50 random
splits of~the~data are reported with standard deviations in
parentheses. For the variable selection only results, the maximum
number of selected wavelengths was restricted to be six to avoid
degeneracies} \label{ta:oliveclassres}
\begin{tabular*}{\textwidth}{@{\extracolsep{\fill}}ld{2.9}@{}}\hline
\textbf{Method} & \multicolumn{1}{c@{}}{\textbf{Misclassification rate}}\\
\hline
Variable selection and updating & 6.9\%\ (5.4)\\
Variable selection (greedy) and updating & 16.6\%\ (11.3)\\
Variable selection only& 17.9\%\ (10.9)\\
\cite{dean06} & 11.9\%\ (6.3)\\
\cite{downey03} & 6.1\%\mbox{--}19.0\%\\
Transductive SVMs & 12.4\%\ (7.5)\\
Random Forests & 19.3\%\ (6.5) \\
AdaBoost.M1 & 34.1\%\ (9.3)\\
Bayesian Multinomial Regression & 57.0\%\ (1.2)\\\hline
\end{tabular*}
\end{table}

A cross tabulation of the classifications with the true origin of
the olive oils (Table~\ref{ta:oliveoilclass}) reveals the difficulty
in classifying the oils.

\begin{table}
\tablewidth=215pt
\caption{Average classification results for the olive oil groups.
The results shown are for the variable selection and~updating~method} \label{ta:oliveoilclass}
\begin{tabular*}{215pt}{@{\extracolsep{\fill}}ld{2.1}d{2.1}d{2.1}@{}}
\hline
&\multicolumn{3}{c@{}}{\textbf{\textit{Predicted}}} \\[-5pt]
&\multicolumn{3}{c@{}}{\hrulefill} \\
\textit{\textbf{Truth}} & \multicolumn{1}{c}{\textbf{Crete}} & \multicolumn{1}{c}{\textbf{Peleponese}}
& \multicolumn{1}{c@{}}{\textbf{Other}}\\
\hline
Crete & 90.0& 8.7& 1.3 \\
Peleponese &1.0 &92.9 & 6.1 \\
Other & 0.0& 3.8& 96.2\\
\hline
\end{tabular*}
\end{table}

\begin{figure}[b]

\includegraphics{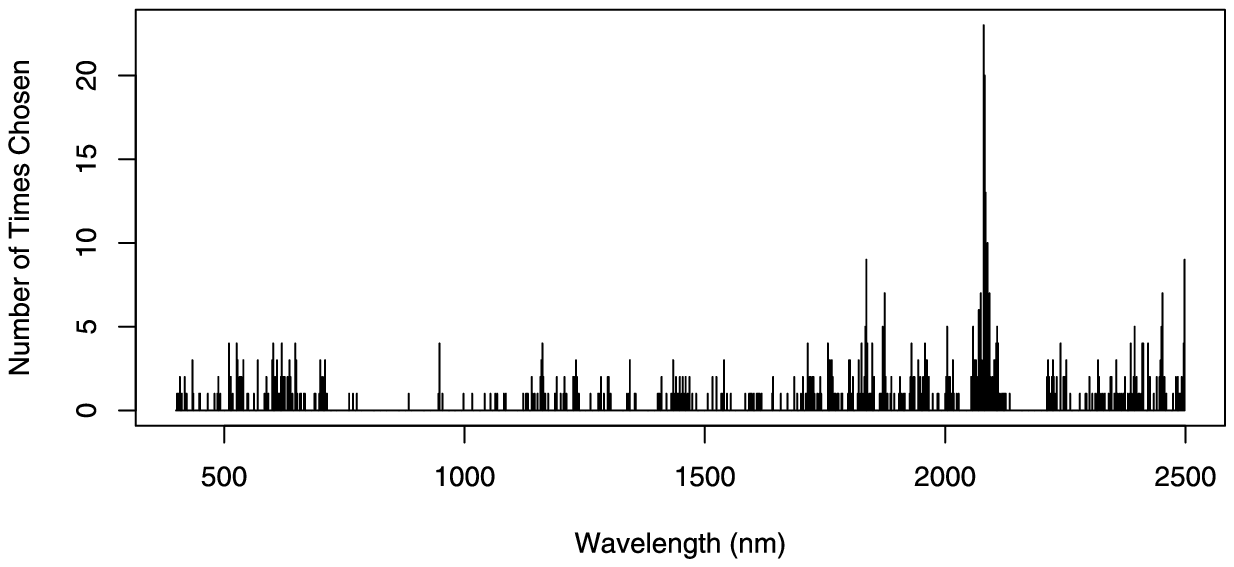}

\caption{Wavelengths chosen in the olive oil classification problem
using the variable selection and updating method. The height of the bars
shows how many times the wavelength was chosen in 50 random splits
of the data.} \label{fi:olivechosen}
\end{figure}

In contrast to the meat classification problem, the chosen
wavelengths for this problem (Figure~\ref{fi:olivechosen}) are
concentrated in the near-infrared region (800--2498~nm), but some
wavelengths in the visible region are also selected. The most
commonly chosen wavelength is 2080~nm, which has been attributed to
an O-H stretching/O-H bend combination [\cite{osborne84}].
Wavelengths near 2310, 2346 and 2386~nm are due to C-H stretching
vibrations and other vibrational modes. In particular, wavelengths
in the 2310~nm region have previously been assigned to fat content.
In all cases, between 6 and 29 wavelengths were selected, with a mean
of~15 wavelengths being chosen. The EEE covariance structure was chosen
for every final model for the olive oil classification problem.

\subsection{Sensitivity to spectral resolution}

In order to determine the sensitivity of the selected wavelengths to
the resolution of the spectrometer used in this study, we investigated
the effect of reducing the number of reflectance values by computing
the mean reflectance value over sets of adjacent wavelengths and using
these as inputs into the variable selection model. The results of this
analysis are outlined for the olive oil authentication problem, and
similar results were found for the meat species authenticity study.

We found that the classification error of the olive oil samples
increases slightly as soon as any adjacent wavelengths are aggregated
(Table~\ref{ta:aggregration}). However, once the wavelengths are
aggregated, the classification error remained steady for aggregating
between 2 and 30 adjacent wavelengths. Thereafter, there was a serious
deterioration in the classification performance when more than 30
adjacent wavelengths were aggregated. This suggests that a considerable
amount of the group information is maintained at even low resolutions,
but that there is more information in the raw data themselves.

\begin{table}
\tablewidth=250pt
\caption{The change in classification performance for the variable
selection and updating method as the number of adjacent
wavelengths~being~aggregated~increases}
\label{ta:aggregration}
\begin{tabular*}{250pt}{@{\extracolsep{\fill}}ld{2.3}@{}}\hline
\textbf{Aggregation level} & \multicolumn{1}{c@{}}{\textbf{Classification error}}\\\hline
\phantom{1}1 & 6.9\%\\
\phantom{1}2 & 9.7\%\\
\phantom{1}3 & 7.6\%\\
\phantom{1}5 & 7.9\%\\
10 & 9.9\%\\
15 & 8.5\%\\
30 & 9.1\%\\
50 & 13.2\%\\
70 & 28.7\%\\
\hline
\end{tabular*}
\end{table}

\begin{figure}

\includegraphics{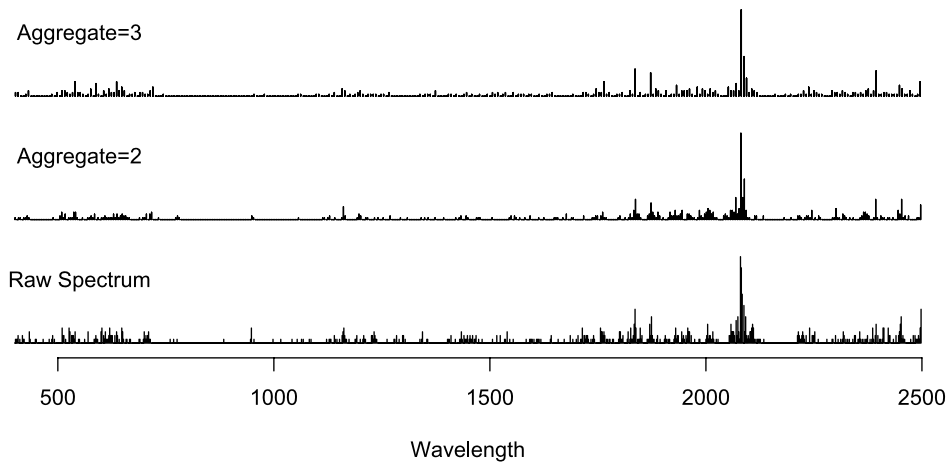}

\caption{The chosen wavelengths when the raw olive oil spectra are
analyzed and when adjacent wavelengths are aggregated.}
\label{fi:aggregration}
\end{figure}

The spectral regions selected when analyzing the data in aggregated
form were found to be stable. In both applications, the selected
regions were very similar for the aggregated data, but fewer variables
tended to be selected because of the aggregation process. Figure~\ref
{fi:aggregration} shows the chosen wavelengths when the raw spectra,
two adjacent wavelengths and three adjacent wavelengths are aggregated
and then analyzed for the olive oil classification problem. This shows
that the selection procedure chooses very specific spectral regions on
both the raw and aggregated scale.

\section{Discussion}
\label{se:conclusions}
The discriminant analysis method presented in this paper gave much better
results than those given by popular statistical and
machine learning techniques such
as Random Forests \citep{breiman01}, AdaBoost \citep{freund97} and
Bayesian Multinomial Regression [\cite{genkin05}, \cite{madigan05}] and
Transductive SVMs [\cite{vapnik95}, \cite{joachims99}] for the
high-dimensional food authenticity data sets analysed here. This
improvement is further enhanced by the addition of the updating
procedure for including the unlabeled data in the estimation method.
The results show that the headlong search
method for variable selection is an efficient method for
selecting wavelengths.

In addition to the improvement in classification results in
the example data sets given, the number of variables
needed for classification was substantially reduced from 1050 to less
than thirty. The variable selection
results in the food authenticity application suggest the possibility
of developing authenticity sensors that only use reflectance values
over a carefully selected subset of the near-infrared and visible
spectral range. The regions of the spectrum selected by the method can
be interpreted in terms of the underlying chemical properties of the
foods under analysis.

We have compared our method with four established leading
classification methods from statistics and machine learning for
which standard software implementations are available. One of these,
AdaBoost, was identified by Leo Breiman as ``the best
off-the-shelf classifier in the world'' [\cite{hastie01}]. It is
possible that the large improvement in performance of our method
relative to the established methods we have compared it with is due to
the fact that our data have many variables of which only a very small
proportion (1\%--3\%) are useful. The variables that are not useful may
introduce a great deal of noise and degrade performance, and so other
methods that do not reduce the number of variables may suffer from this.

Although the methods were developed for the food authenticity
application outlined herein, the method could be applied in contexts
such as the analysis of gene expression data and
document classification. The results of the variable selection
procedure could mean a substantial savings in
terms of time for data collection and space for future data storage.

A range of recent approaches to variable selection in a
classification context include the DALASS approach of
\cite{trendafilov07}, variable selection for kernel Fisher
discriminant analysis [\cite{louw06}] and the stepwise stopping rule
approach of \cite{munita06}. A number of different search algorithms
(proposed as alternatives to backward/forward/stepwise search)
wrapped around different discriminant functions are compared by
\cite{pacheco06}, and genetic search algorithms wrapped around Fisher
discriminant analysis are considered by \cite{chiang04}. Another
example of variable selection methods in the
context of classification using spectroscopic data is given by \cite{indahl04}.

In terms of other approaches to variable selection, a good review
of recent work on the problem of variable or feature
selection in classification was given by \cite{guyon03} from a
machine learning perspective. A good
review of methods involving Support Vector Machines (SVMs) (along
with a proposed criterion for exhaustive variable selection) is
given by \cite{maryhuard07}. An extension allowing variable
selection for the multiclass problem using SVMs is given by
\cite{wang07}. An alternative approach for combining pairwise
classifiers, based on \cite{hastie98}, is given by \cite{szepannek06}.
\cite{greenshtein06} looks at theoretical aspects of
the $n \ll p$ classification and variable selection problem in terms
of empirical risk minimization subject to $l_1$ constraints. Finally, an
alternative to single subset variable selection through Bayesian
Model Averaging [\cite{madigan94}] is given by \cite{dash04}.

\section*{Acknowledgments}

We would like to thank the Editor, Associate Editor and Referees whose
suggestions greatly improved this paper. We would also like to thank
Gerard Downey for providing the food authenticity data and for help
with interpreting the results of the analysis.

\begin{supplement}[id=suppA]
\sname{Supplement}
\stitle{Data}
\slink[doi]{10.1214/09-AOAS279SUPP}
\slink[url]{http://lib.stat.cmu.edu/aoas/279/supplement.zip}
\sdatatype{.zip}
\sdescription{This zipfile [Murphy, Dean and Raftery (\citeyear{murphy09})] contains the data sets used in this paper.
The original data source information and conditions for the use of the
data are outlined in this file.}
\end{supplement}

\printaddresses

\end{document}